\pdfoutput=1

\documentclass[11pt]{article}

\usepackage{tabularx}
\usepackage{amsmath}
\usepackage{amssymb}
\usepackage{graphicx}
\usepackage{subfigure}
\usepackage{multirow}
\usepackage{algorithm}  
\usepackage{algpseudocode}  
\usepackage[table]{xcolor}
\usepackage{makecell}
\usepackage{arydshln}

\usepackage[final]{coling}

\usepackage{times}
\usepackage{latexsym}

\usepackage[T1]{fontenc}

\usepackage[utf8]{inputenc}

\usepackage{microtype}

\usepackage{inconsolata}

\usepackage{graphicx}

\title{Towards Real-World Rumor Detection: Anomaly Detection Framework with Graph Supervised Contrastive Learning}

\author{
 \textbf{Chaoqun Cui},
 \textbf{Caiyan Jia\thanks{Corresponding author.}}
\\
\\
School of Computer Science and Technology \& Beijing Key Lab of Traffic Data\\ Analysis and Mining Beijing Jiaotong University, Beijing 100044, China
\\
 \texttt{ccqun19990728@gmail.com}\\
 \texttt{cyjia@bjtu.edu.cn}
}

\begin{document}
\maketitle
\begin{abstract}

Current rumor detection methods based on propagation structure learning predominately treat rumor detection as a class-balanced classification task on limited labeled data. However, real-world social media data exhibits an imbalanced distribution with a minority of rumors among massive regular posts. To address the data scarcity and imbalance issues, we construct two large-scale conversation datasets from Weibo and Twitter and analyze the domain distributions. We find obvious differences between rumor and non-rumor distributions, with non-rumors mostly in entertainment domains while rumors concentrate in news, indicating the conformity of rumor detection to an anomaly detection paradigm. Correspondingly, we propose the Anomaly Detection framework with Graph Supervised Contrastive Learning (AD-GSCL). It heuristically treats unlabeled data as non-rumors and adapts graph contrastive learning for rumor detection. Extensive experiments demonstrate AD-GSCL's superiority under class-balanced, imbalanced, and few-shot conditions. Our findings provide valuable insights for real-world rumor detection featuring imbalanced data distributions.

\end{abstract}

\section{Introduction}

Numerous studies \cite{bigcn,ragcl} highlight the value of propagation structures in revealing inter-post relations for rumor detection. Datasets like Weibo \cite{weibo}, DRWeibo \cite{ragcl}, Twitter15, Twitter16 \cite{twitter1516}, and PHEME \cite{pheme} are utilized, containing source posts and replies. Propagation structure based methods convert a claim into a tree structure, with source posts as roots, replies as nodes, and reply relations as edges (see Appendix~\ref{sec:rpt} for examples). Graph Neural Networks (GNNs) then learn this tree to detect authenticity.

Contemporary rumor detection methods based on propagation structure learning typically treat rumor detection as a classification task, predominantly run on small-scale, class-balanced datasets. However, these techniques encounter several prevailing challenges: (1) Owing to effective post-screening measures conducted by social media platforms, numerous rumor posts are promptly identified and eradicated upon their publication. This results in difficulties in securing a sufficiently large corpus of labeled data; (2) Practically, rumor detection on social media platforms inherently entails the identification of a relatively minute quantity of rumor posts within a massive pool of posts, the vast majority of which are non-rumors. Therefore, the extant paradigm, founded on class-balanced classification tasks, lacks alignment with the genuine application scenarios; (3) Posts circulating on social media platforms traverse multiple diverse domains, each showcasing unique characteristics. Consequently, the data cross-domain nature introduces added complexities to the task. Our ambition is to address these three predominant issues.

In this study, we initially construct two large-scale unlabeled conversation datasets from Weibo and Twitter using web crawler, aligning with real-world data distributions. We then investigate the domain distribution of data on these platforms. Our findings lead us to heuristically frame rumor detection as an anomaly detection task rather than a conventional class-balanced classification task, more accurately reflecting real-world scenarios. Specifically, we propose the \textbf{A}nomaly \textbf{D}etection framework with \textbf{G}raph \textbf{S}upervised \textbf{C}ontrastive \textbf{L}earning (AD-GSCL). Guided by our findings, AD-GSCL treats large-scale unlabeled data as non-rumor and smartly upgrades existing graph contrastive learning methods \cite{scl,infograph} to adapt to rumor detection's domain distribution.

In summary, this study contributes as follows:
\begin{itemize}
\item We built two large-scale unlabeled conversation datasets to mitigate labeled data scarcity.
\item We analyze social media's data domain distribution, highlighting significant discrepancies between rumor and non-rumor class data.
\item We propose AD-GSCL, which employs an anomaly detection framework to align with real-world scenarios.
\item Experiments demonstrate AD-GSCL's superior performance in multiple scenarios.
\end{itemize}

\section{Related Work}

In this section, we review the related works on rumor detection and graph contrastive learning.

\subsection{Rumor Detection}

Early studies in rumor detection used traditional classification with hand-crafted features \cite{dtc,feature2,feature3}. Deep learning's success has spawned numerous methods that significantly enhance rumor detection performance. These methods can be roughly divided into four classes, including time-series based methods \cite{yucnn,user3,user4} which model text content or user profiles as time series, propagation structure learning methods \cite{rvnn,treetrans,ragcl,debiased,pep,urumor} which consider the propagation structures of source claims and their replies, multi-source integation methods \cite{ms1,ms3,ms4,ms2} which combine multiple resources of rumors including post content, user profiles, heterogeneous relations between posts and users, multi-modal fusion methods \cite{otherrumor1,mm2,mm3,vga} which use posts' content and their related images to debunk rumors. 

Although large language models have achieved remarkable performance in many natural language processing tasks \cite{gpt3,rlhf,sspo}, propagation structure information remains crucial in rumor detection. Numerous state-of-the-art (SOTA) models \cite{clahi,gacl} bank on learning the representations of rumor propagation trees utilizing GNNs. 
Traditional approaches treat rumor detection as a class-balanced task, diverging from real-world scenarios. Recent studies recognize the inherent data imbalance in practical settings and use large volumes of unlabeled data through anomaly detection or Positive-Unlabeled (P-U) learning frameworks.
For example, \citet{adrd2} unveiled a multi-modal approach that was capable of incrementally processing platform data streams. \citet{adrd3} formulated a detection schema predicated on anomaly signal scoring. \citet{pu2,pu1} leveraged P-U learning on heterogeneous networks. Moreover, \citet{adrd1} introduced an efficacious streaming framework to comply with the latency bounds inherent in the real-time processing of large-scale social media data streams. Nonetheless, these methods neglected to effectively employ the propagation structure from user feedback (i.e., comments). 
In light of the above considerations, AD-GSCL tackles imbalanced data distribution with anomaly detection paradigm, learning from claim propagation structure information via GNNs, and adapting SOTA graph contrastive learning for rumor detection.

\subsection{Graph Contrastive Learning}

Deep learning advancements have led to progress in neural message passing algorithms \cite{messagepass}, which set new SOTA benchmarks in various tasks \cite{messagepass1,messagepass2,messagepass3} through supervised learning of graph representations. However, the challenge of acquiring labeled data has shifted focus to graph self-supervised learning methods \cite{mvgrl,gca,graphmae}, which utilize unlabeled data to learn robust node or graph representations. Most of them are graph contrastive learning methods including methods based on mutual information (MI) maximization \cite{dgi,infograph} and graph augmentation \cite{graphcl,joao}. 
The former are usually trained by maximizing the MI between local representations and global representation. 
The latter firstly use different graph augmentation strategies to obtain different views of a given graph, and then construct positive and negative samples. Finally, graph representations are learned by minimizing the contrastive loss.
These methods provide powerful ways to extract the discriminative features from rumor propagation trees.
In order to adapt to different types of graph datasets, there are also some researches on adaptive graph augmentation \cite{gca,joao,autogcl}, which are able to automatically select optimal augmentation strategies for different datasets. 
AD-GSCL does not necessitate the design of intricate network architectures. Rather, it combines the existing supervised contrastive learning \cite{scl} and unsupervised contrastive learning \cite{infograph} paradigms, and innovates upon these frameworks in alignment with the domain distribution of the datasets. 

\section{Analysis on Domain Distribution}

Social media platforms, striving for a healthy network environment, regulate posts via automated and manual methods. The prompt removal of detected rumors results in a scarcity of rumor data, reflecting in the limited size of labeled datasets (Table~\ref{tab:sta}). In our study, we reviewed the most recent 500 user-reported rumors displayed on Weibo rumor-refutation platform\footnote{\url{https://service.account.weibo.com/?type=5}} from July 29, 2023, to April 5, 2024. Out of these, only 84 corresponding source posts remained undeleted. This indicates that social platforms like Weibo are very stringent in the control of rumors and misinformation.

For the issue of scarce labeled data, unsupervised or semi-supervised methods serve as apt solutions, as they harness massive unlabeled data to learn unsupervised representations. Several studies have embarked on rumor detection leveraging large-scale data streams from social media platforms for semi-supervised rumor detection \cite{adrd2,adrd3,adrd1}, however, such streams typically do not include the propagation structure information. 
We utilized web crawlers to create two large-scale unlabeled datasets from Weibo and Twitter, termed Unlabeled Weibo (UWeibo) and Unlabeled Twitter (UTwitter). These datasets, derived from trending posts on both platforms, include complete claim propagation structures. Dataset construction details are in Appendix~\ref{sec:udc}, and Table~\ref{tab:sta} presents the statistics of the datasets harnessed in our study.

\begin{table*}[h]
\centering
\resizebox{\textwidth}{!}{
\begin{tabular}{cccccccc}
\Xhline{1.0pt}
\rowcolor{gray!20}
\textbf{Statistic} & \textbf{Weibo} & \textbf{DRWeibo} & \textbf{Twitter15} & \textbf{Twitter16} & \textbf{PHEME} & \textbf{UWeibo} & \textbf{UTwitter}\\
\hline
\textbf{language} & zh & zh & en & en & en & zh & en \\
\textbf{labeled} & True & True & True & True & True & False & False \\
\hdashline
\textbf{\# claims} & 4664 & 6037 & 1490 & 818 & 6425 & 209549 & 204922\\
\textbf{\# non-rumors} & 2351 & 3185 & 374 & 205 & 4023 & - & -\\
\textbf{\# false rumors} & 2313 & 2852 & 370 & 205 & 638 & - & -\\
\textbf{\# true rumors} & - & - & 372 & 207 & 1067 & - & -\\
\textbf{\# unverified rumors} & - & - & 374 & 201 & 697 & - & -\\
\hdashline
\textbf{\# avg reply} & 803.5 & 61.8 & 50.2 & 49.1 & 15.4 & 45.0 & 83.9\\
\textbf{\# avg shallow ($\leq 2$) reply} & 692.2(86\%) & 59.1(95\%) & 41.4(83\%) & 37.6(76\%) & 10.8(70\%) & 46.6(92\%) & 70.0(85\%)\\
\textbf{\# avg deep ($> 2$) reply} & 111.2(14\%) & 2.7(5\%) & 8.7(17\%) & 11.5(24\%) & 4.6(30\%) & 4.0(8\%) & 12.5(15\%)\\
\Xhline{1.0pt}
\end{tabular}
}
\caption{Statistics on the distribution of dataset classes and node depth.}
\label{tab:sta}
\end{table*}

The intricate nature of the real world ensures that social posts cross multiple domains (like sport, science, culture, etc.), with substantial disparity in attributes and content structures among these domains \cite{crossdomain1,crossdomain3}. The inability of rumor detection models to efficaciously adapt to these domain-specific differences will markedly influence their performance. We utilize two unlabeled datasets and several labeled datasets to investigate the domain distributions of social media data and varying data classes. Specifically, we further divide major social media domains into three sections: 
(1) \textbf{Entertainment Section}: This includes domains unlikely to propagate misinformation or rumors, like personal life sharing, sports. Most social media domains reside within this section. These domains typically contain less sensitive content with minimal public opinion impact. Their authenticity is usually non-controversial.
(2) \textbf{Knowledge Section}: This covers domains disseminating knowledge across various fields like popular science, healthcare. Posts here often share specific knowledge, sometimes containing misinformation or falsified details.
(3) \textbf{News Section}: This includes domains focusing on news content (e.g., social, financial news). These posts are often news-oriented, occasionally containing rumors.

\begin{figure*}[h]
  \centering
  \subfigure[Domain]{\includegraphics[width=0.766\textwidth]{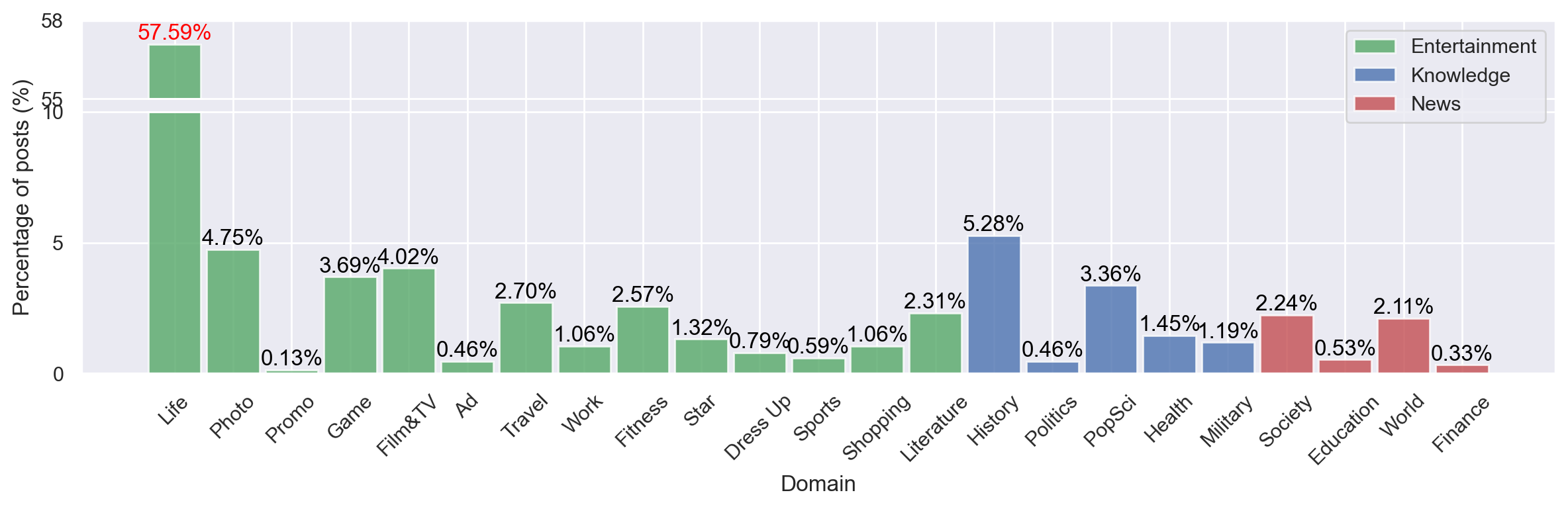}}
  \subfigure[Section]{\includegraphics[width=0.184\textwidth]{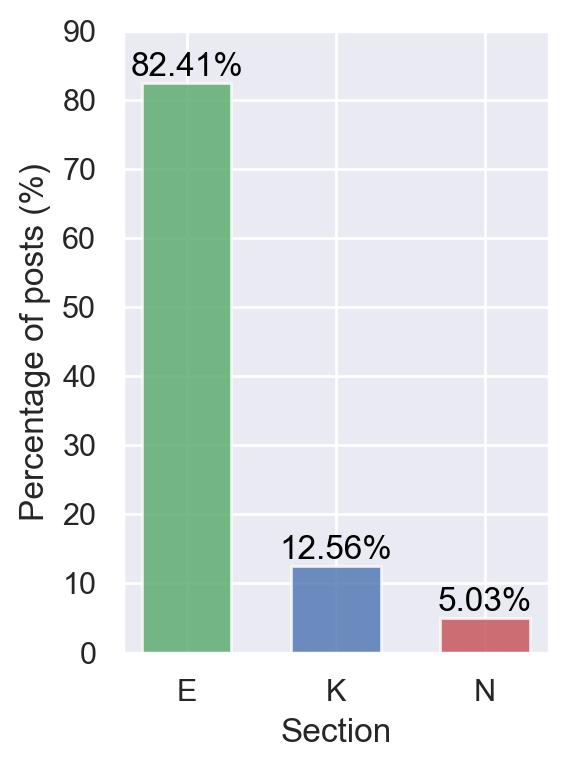}}
  \caption{Domain and section distribution of posts on UWeibo dataset. E, K, and N represent the three sections.}
  \label{fig:ds_uweibo}
\end{figure*}

\begin{figure}[h]
  \centering
  \subfigure[Non-Rumor]{\includegraphics[width=0.184\textwidth]{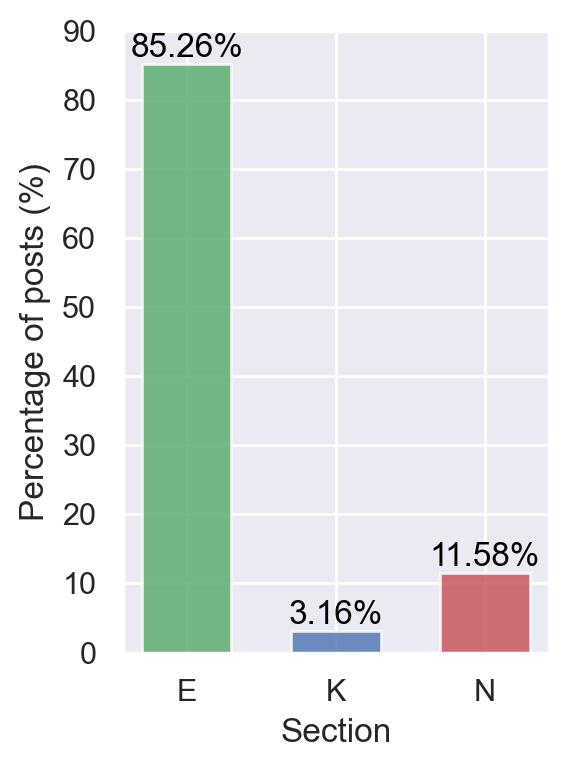}}
  \subfigure[Rumor]{\includegraphics[width=0.184\textwidth]{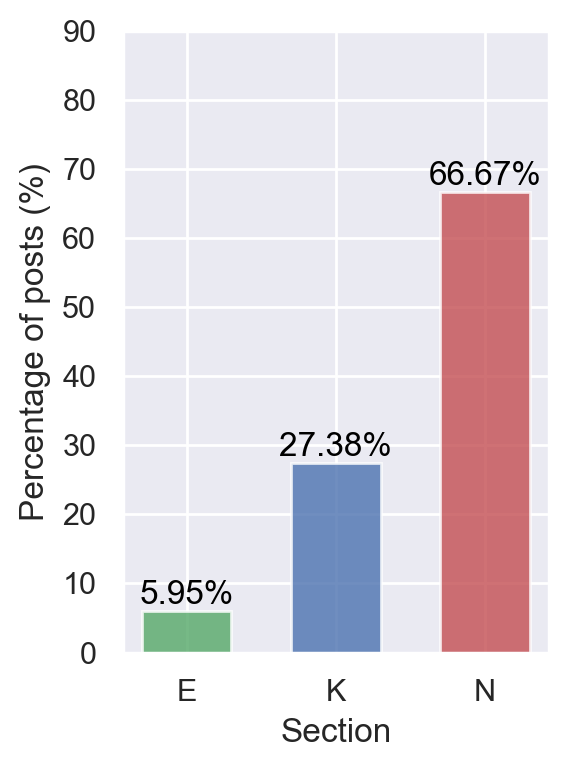}}
  \caption{Post section distribution on Weibo dataset.}
  \label{fig:section_weibo}
\end{figure}

For Weibo platform data, we ran web scraping to extract domain labels, enabling us to statistically analyze the domain and section distribution. 
Our statistical analysis of the domain and section distribution of UWeibo dataset is presented in Figure~\ref{fig:ds_uweibo}, providing an insight into the typical content structure of trending posts on Weibo platform. We also scrutinize the section distribution of non-rumor and rumor classes within Weibo dataset, as depicted in Figure~\ref{fig:section_weibo}. From this analysis, we discern that: (1) A significant majority of posts in UWeibo reside in Entertainment section (82.41\%), with a substantial proportion of these being personal life sharing content (57.79\%), which is also consistent with the main purpose of social media platforms; (2) A marked discrepancy is observed in the section distribution between the non-rumor and rumor classes in Weibo dataset, with the non-rumor class predominantly in Entertainment section (85.26\%), whereas the Rumor class primarily features in News section (66.67\%); (3) The section distribution of unlabeled data in UWeibo is very similar to the non-rumor in Weibo, i.e. most posts belong to Entertainment section that is less likely to generate rumors.

Informed by the aforementioned survey and some intuitive considerations, we determined a pivotal configuration in AD-GSCL, namely, the heuristic treating of unlabeled data from UWeibo and UTwitter as the non-rumor class. We justify this decision based on: (1) Intuitively, social media platforms monitor and regulate posts from users, so rumor posts can be deleted after being detected, thus rarely appearing among trending posts; (2) A considerable portion of trending platform posts often come from high-follower-count users, thus their credibility is ensured to some extent; (3) The majority of UWeibo belongs to Entertainment section that is less likely to generate rumors, and its section distribution is highly consistent with the non-rumor class. Moreover, data may be classified in a more fine-grained manner, such as Non-Rumor, True Rumor, False Rumor, and Unverified Rumor. To substantiate whether unlabeled data can be assigned to classes other than non-rumor, we also examined the data distribution of UTwitter and the section distributions of different classes in Twitter15 and Twitter16 (see Appendix~\ref{sec:dsm}). This analysis reveals a similar section distribution in UTwitter to UWeibo (with Entertainment section posts being predominant), and that the section distribution of the non-rumor class significantly diverges from the other three rumor classes in Twitter15 and Twitter16, while being highly consistent with the UTwitter section distribution. These auxiliary findings underscore the broader applicability of our survey results across diverse platforms and support the rationale of designating unlabeled data as non-rumor. By adopting this configuration, AD-GSCL frames rumor detection within an anomaly detection paradigm—identifying the minority of rumor data (anomalous samples) within the majority of non-rumor data (normal samples). 

\section{Method}

In this section, we present the design of AD-GSCL.

\subsection{Notation}

Rumor detection is a graph classification task whose goal is to predict the class of labeled claims with conversation context. Specifically, we denote the labeled claim set as $\mathbb{C}^{L}=\left \{c_{1},c_{2},\cdots ,c_{N_{L}}\right \}$. Each claim $c=(y,G)$ consists of its ground truth label $y\in \left \{N,R\right \}$ (i.e., Non-Rumor or Rumor) or fine-grained label $y\in \left \{N,F,T,U\right \}$ (i.e., Non-Rumor, False Rumor, True Rumor, Unverified Rumor) and its conversation context modeling as a propagation graph $G=(V,E)$, where $V$ and $E$ represent the set of nodes and edges, respectively. The set of propagation graphs corresponding to labeled claims is $\mathbb{G}^{L}=\left \{G_{1},G_{2},\cdots ,G_{N_{L}}\right \}$. 
The initial feature matrix corresponding to $V$ is $X\in \mathbb{R}^{|V|\times d_{x}}$, and $d_{x}$ is the feature vector dimension. The initial feature of node can be selected widely such as Word2vec \cite{word2vec}, tf-idf features, or vectors extracted from pre-trained language models \cite{roberta,electra}. 
In addition, we denote the unlabeled claims set as $\mathbb{C}^{U}=\left \{c_{N_{L}+1},c_{N_{L}+2},\cdots ,c_{N_{L}+N_{U}}\right \}$. 
There is no corresponding $y$ for unlabeled claims, but only the propagation structure $G$ for each claim. 
The propagation graph set corresponding to unlabeled claims is $\mathbb{G}^{U}=\left \{G_{N_{L}+1},G_{N_{L}+2},\cdots ,G_{N_{L}+N_{U}}\right \}$. Our goal is to learn high-quality representations of graphs in $\mathbb{G}^{L}$ and $\mathbb{G}^{U}$, simultaneously train a classifier to predict the class of claims in $\mathbb{C}^{L}$.

\subsection{Framework}

\begin{figure*}[h]
  \centering
  \includegraphics[width=\textwidth]{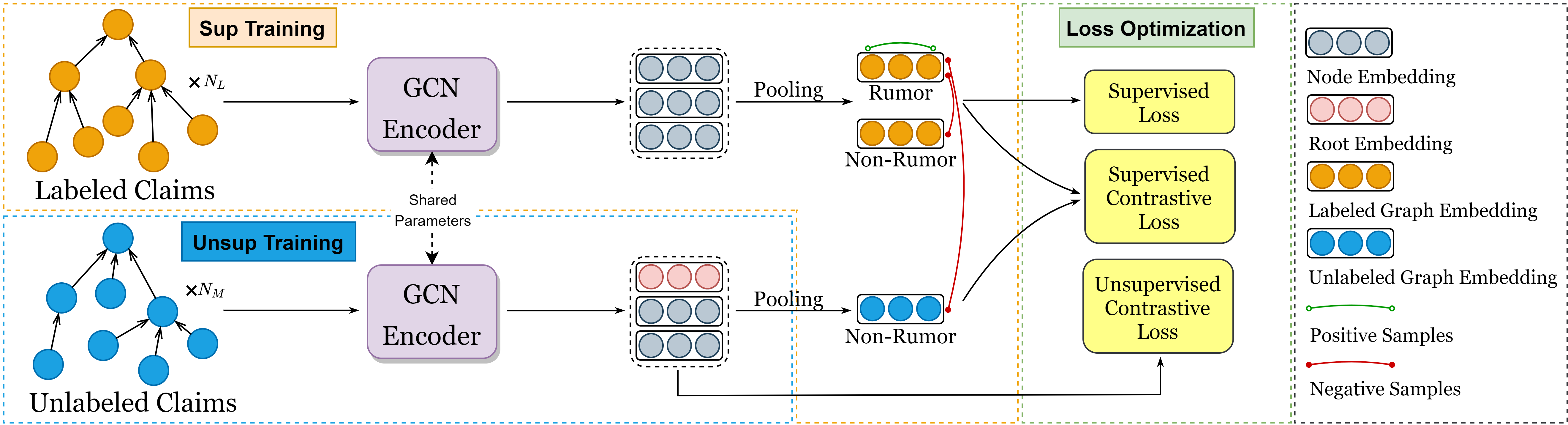}
  \caption{The framework of AD-GSCL.}
  \label{fig:framework}
\end{figure*}

AD-GSCL solely requires modification and integrations of existing contrastive losses, bypassing the need for intricate architectural designs. All claims within the unlabeled dataset are treated as non-rumor instances. Given the cross-domain nature of non-rumor claims, we refrain from constructing positive samples between non-rumor claims within the supervised contrastive loss. Hence, to secure a good representation of non-rumor instances, we employ unsupervised contrastive loss to learn the non-rumor samples within our unlabeled dataset. The framework of AD-GSCL is shown in Figure~\ref{fig:framework}.

\subsection{Supervised Contrastive Learning}

In AD-GSCL, rumor data is sourced from labeled benchmark datasets, whereas non-rumor data derives from both labeled and unlabeled datasets. AD-GSCL employs supervised contrastive learning to enhance the inter-class representation dissimilarity and intra-class representation similarity of different class claims in the feature space. Specifically, we designate the claims within each class, excluding those from non-rumor classes, as positive samples for each other, whereas claims spanning different classes are treated as negative samples. Notably, non-rumor class claims are not deemed as positive samples. The rationale behind this methodology of positive sample construction is backed by our survey findings depicted in Figure~\ref{fig:ds_uweibo}. The UWeibo data mainly belongs to Entertainment section. The vast majority of domains in social media belong to this section, indicating that the UWeibo data has very broad cross-domain characteristics. Consequently, it would be unreasonable to pull these data with severe cross-domain issues closer together as positive samples in the feature space. In contrast, rumor claims tend to mainly fall under News section and to a lesser extent, Knowledge section. These sections typically feature a smaller number of domains, thus implying a lesser degree of cross-domain characteristics. Moreover, user replies to rumor claims frequently exhibit stronger stances and emotional expressions \cite{rvnn,ebgcn}, thus yielding more prominent domain invariant features. Consequently, the construction of positive samples for the supervised contrastive loss in AD-GSCL relies solely on these rumor claims. 

Specifically, for labeled dataset $\mathbb{C}^{L}$ and unlabeled dataset $\mathbb{C}^{U}$, the graph representation of a propagation graph $G\in \mathbb{G}^{L}\cup \mathbb{G}^{U}$ obtained by GCN encoder \cite{gcn} is denoted as $h_{G}$. The supervised contrastive loss is
\begin{small}
\begin{equation}
\begin{split}
&\mathcal{L}_{scl}(\mathbb{C}^{L}\cup \mathbb{C}^{U})=\\
&-\frac{1}{|\mathbb{G}_{R}^{L}|}\sum _{G\in \mathbb{G}_{R}^{L}}log\left \{\frac{1}{|P(G)|}\frac{\sum _{G_{p}}exp(sim(h_{G},h_{G_{p}})\tau )}{\sum _{G_{n}}exp(sim(h_{G},h_{G_{n}})\tau )}\right \},
\end{split}
\end{equation}
\end{small}
where $\mathbb{G}_{R}^{L}$ is the set of rumor class claim propagation graphs in dataset $\mathbb{C}^{L}$; $p\in P(G)$ and $n\in N(G)$ are indices of positive and negative samples respectively; $P(G)\equiv \left \{p:G_{p}\in \mathbb{G}^{L},y_{G_{p}}=y_{G}\right \}$ and $N(G)\equiv \left \{n:G_{n}\in \mathbb{G}^{L}\cup \mathbb{G}^{U},y_{G_{n}}\neq y_{G}\right \}$ are the sets of positive and negative sample indices respectively; $sim(\cdot )$ is the cosine similarity; $\tau \in \mathbb{R}^{+}$ is a scalar temperature parameter. 

\subsection{Anchor Representation Learning}

Supervised contrastive learning primarily aims to use massive non-rumor claims (normal samples) as anchor samples, reducing the similarity between the representations of rumor claims (anomalous samples) and anchor samples in feature space, thereby aiding rumor classification. Consequently, learning robust representations for anchor samples is crucial.
We previously asserted that increasing the representation similarity among non-rumor claims by considering them as mutual positive samples in supervised contrastive learning may not be logical, given the diverse domains these claims may originate from. Consequently, we resort to unsupervised contrastive learning in AD-GSCL to learn the representations of anchor samples. In particular, a graph contrastive learning approach based on MI maximization \cite{infograph} has been employed. The positive samples of unsupervised contrastive learning are derived from both global and local representations of an individual instance, ensuring the absence of amplified representation similarity between claims of divergent domains. As reported in Table~\ref{tab:sta}, most nodes are located at shallow positions in the propagation tree of claims in social media, i.e., they are very close to the root node. During the forward propagation in GNN, nodes aggregate their k-hop neighborhood nodes features \cite{messagepass,gcn}. In AD-GSCL, we employ the bottom-up propagation tree, so the root node can aggregate the features of most nodes in the graph. Therefore, we directly use the root representation as the global representation to compute the MI maximization contrastive loss with node representations, which is another specific design in AD-GSCL. Since it is crucial for rumor detection to model the interaction between source posts and replies \cite{bigcn,gacl}, this design that emphasizes the importance of the root node is more suitable for this task.

Specifically, for a claim in $\mathbb{C}^{U}$ with propagation tree $G=(V,E)$, the node representations and the corresponding root representation obtained from the encoder are denoted as $h_{v}$ ($v\in V$) and $h_r$. We use Jensen-Shannon MI estimator \cite{js} to calculate the MI as below.

\begin{small}
\begin{equation}
\begin{split}
 I(h_{v}(G);h_{G}):=\mathbb{E}_{\mathbb{P}}[-sp(-T(h_{v}(G),h_{r}(G))]\\-\mathbb{E}_{\mathbb{P}\times \tilde{\mathbb{P}}}[sp(T(h_{v}(G^{'}),h_{r}(G)))],
\end{split}
\end{equation}
\end{small}
where $\mathbb{P}$ is the distribution followed by $\mathbb{G}^{U}$; $G$ is an input graph sampled from $\mathbb{P}$; $G^{'}$ is a negative sample sampled from $\tilde{\mathbb{P}}=\mathbb{P}$; $T$ is a discriminator; $sp(z)=log(1+e^{z})$ is the softplus function. In practice, we use all combinations of root representations and node representations of all graph instances within a batch to generate negative samples. The unsupervised contrastive loss on $\mathbb{C}^{U}$ is
\begin{small}
\begin{equation}
\mathcal{L}_{ucl}(\mathbb{C}^{U})=-\frac{1}{N_{U}}\sum _{G\in \mathbb{G}^{U}}\sum _{v\in V}I(h_{v}(G);h_{r}(G)).
\end{equation}
\end{small}

\subsection{Training Strategy}

AD-GSCL can be trained with pre-training strategy and semi-supervised strategy. Different training strategies will minimize $\mathcal{L}_{sup}(\mathbb{C}^{L})$, $\mathcal{L}_{scl}(\mathbb{C}^{L}\cup \mathbb{C}^{U})$ and $\mathcal{L}_{ucl}(\mathbb{C}^{U})$ in different ways.

\subsubsection{Pre-Training Strategy}

AD-GSCL can be pretrained on unlabeled dataset $\mathbb{C}^{U}$ to minimize $\mathcal{L}_{ucl}(\mathbb{C}^{U})$. The process is aimed at learning more robust representations of non-rumor claims so that they can better serve as anchors:
\begin{equation}
\mathcal{L}_{pre}=\mathcal{L}_{ucl}(\mathbb{C}^{U}).
\end{equation}

AD-GSCL will continue to minimize $\mathcal{L}_{scl}(\mathbb{C}^{L}\cup \mathbb{C}^{U})$ after pre-training to learn discriminative features of different class claims, and also minimize cross-entropy classification loss $\mathcal{L}_{sup}(\mathbb{C}^{L})$, in which $\mathcal{L}_{scl}(\mathbb{C}^{L}\cup \mathbb{C}^{U})$ is a regularization term. This process is similar to fine-tuning, whose loss is
\begin{equation}
\mathcal{L}_{ft}=\mathcal{L}_{sup}(\mathbb{C}^{L})+\lambda \cdot \mathcal{L}_{scl}(\mathbb{C}^{L}\cup \mathbb{C}^{U}),
\end{equation}
where $\lambda$ is an adjustable hyperparameter. The whole process is shown in Algorithm~\ref{alg:pre}.

\begin{algorithm}[t]
  \caption{Pre-training Strategy Optimization}
  \label{alg:pre}
  \begin{algorithmic}[1]
    \Require initial model parameter $\theta ^{(0)}$, pre-training step $N$, fine-tuning step $M$.
    \Ensure optimized model parameter $\theta ^{(N+M)}$.
    \State $//$ Pre-training on $\mathbb{C}^{U}$. 
    \For {$n=1$ to $N$}
    \State Minimize $\mathcal{L}_{pre}$ to update $\theta ^{(n)}$.
    \EndFor
    \State $//$ Fine-tuning on $\mathbb{C}^{L}$ and $\mathbb{C}^{U}$. 
    \For {$m=1$ to M}
    \State Minimize $\mathcal{L}_{ft}$ to update $\theta ^{(N+m)}$.
    \EndFor\\
    \Return $\theta ^{(N+M)}$.
  \end{algorithmic}
\end{algorithm}

\subsubsection{Semi-Supervised Strategy}

AD-GSCL uses the combination of all the loss functions for semi-supervised learning. In this case, $\mathcal{L}_{sup}(\mathbb{C}^{L})$ is the main loss, $\mathcal{L}_{scl}(\mathbb{C}^{L}\cup \mathbb{C}^{U})$ and $\mathcal{L}_{ucl}(\mathbb{C}^{U})$ are regularization terms, then the loss function of semi-supervised training strategy is
\begin{equation}
\begin{split}
\mathcal{L}_{total}=\mathcal{L}_{sup}(\mathbb{C}^{L})+\alpha \cdot \mathcal{L}_{scl}(\mathbb{C}^{L}\cup \mathbb{C}^{U})\\+\beta \cdot \mathcal{L}_{ucl}(\mathbb{C}^{U}),
\end{split}
\end{equation}
where $\alpha$ and $\beta$ are adjustable hyperparameters.

\section{Experiments}

We present main experiments in this section. See Appendix~\ref{sec:se} for further Evaluations.

\subsection{Experimental Settings}

We conducted experiments on datasets Weibo, DRWeibo, Twitter15, Twitter16, and PHEME. 
Table~\ref{tab:sta} shows the statistics of all the datasets. UWeibo and UTwitter are available at \url{https://github.com/CcQunResearch/UWeibo} and \url{https://github.com/CcQunResearch/UTwitter}.

We compare with the following baselines: 

\textbf{PLAN} \cite{plan} uses Transformer architecture to detect rumors.

\textbf{BiGCN} \cite{bigcn} leverages two bidirectional GCN encoders and root node feature enhancement strategy.

\textbf{GACL} \cite{gacl} utilizes contrastive learning and adversarial training to classify rumor.

\textbf{DDGCN} \cite{ddgcn} can model multiple types of information in one unified framework.

\textbf{RAGCL} \cite{ragcl} uses contrastive learning with adaptive data augmentation.

Experiment setting details are in Appendix~\ref{sec:ed}. Weibo and DRWeibo datasets are used alongside UWeibo, and Twitter15, Twitter16, and PHEME are paired with UTwitter. The code is at \url{https://github.com/CcQunResearch/AD-GSCL}.

\subsection{Class-Balanced Classification}

We conducted classification experiments on class-balanced datasets of Weibo, DRWeibo, Twitter15, and Twitter16, maintaining class-balanced test sets. Average results from 10 random splits of labeled datasets are reported in Table~\ref{tab:weibo-cbc} and ~\ref{tab:twitter-cbc}. We evaluated AD-GSCL's performance under two training strategies: pre-training (`p') and semi-supervised training (`s'). AD-GSCL outperformed the compared baseline methods across all datasets, achieving an accuracy improvement of 1.2-2.8\%. PLAN, using the Transformer architecture, underperform and consume significant GPU resources, highlighting the need for GNN architecture. While GACL, utilizing BERT \cite{bert} for initial feature extraction, exhibits negligible improvement over other baselines. This observation suggests that the mechanism of initial feature extraction might be less critical to rumor detection models compared to the high-level model's capacity to learn node interactions. Traditional methods using GNNs and contrastive learning face issues like overfitting, over-smoothing, and cross-domain discrepancies on limited-scale datasets. In stark contrast, AD-GSCL utilizes large-scale unlabeled data to mitigate the limitations of small data scale, adjusts the data distribution to align with real-world scenarios, and addresses cross-domain discrepancies, thus contributing to improved performance. 

\begin{table*}[h]
\centering
\begin{tabular}{cccccccccc}
 \Xhline{1.0pt}
 \rowcolor{gray!20}
 ~ & ~ & \multicolumn{4}{c}{\textbf{Weibo}} & \multicolumn{4}{c}{\textbf{DRWeibo}}\\
 \cline{3-10}
 \rowcolor{gray!20}
 \multirow{-2}{*}{\textbf{Method}} & \multirow{-2}{*}{\textbf{Class}} & \textbf{Acc.} & \textbf{Prec.} & \textbf{Rec.} & \textbf{F1} & \textbf{Acc.} & \textbf{Prec.} & \textbf{Rec.} & \textbf{F1}\\
 \hline
 \multirow{2}{*}{PLAN} & R & \multirow{2}{*}{91.5\small ±0.7} & 90.8 & 92.3 & 91.5 & \multirow{2}{*}{78.8\small ±0.5} & 78.6 & 76.0 & 77.1 \\
 & N & ~ & 92.3 & 90.7 & 91.4 & ~ & 79.3 & 81.3 & 80.2 \\
 \hline
 \multirow{2}{*}{BiGCN} & R & \multirow{2}{*}{94.2\small ±0.8} & 91.9 & 96.8 & 94.2 & \multirow{2}{*}{86.6\small ±1.0} & 86.9 & 84.9 & 85.8 \\
 & N & ~ & 96.7 & 91.8 & 94.2 & ~ & 86.3 & 88.2 & 87.2 \\
 \hline
 \multirow{2}{*}{GACL} & R & \multirow{2}{*}{93.8\small ±0.6} & 93.6 & 94.0 & 93.8 & \multirow{2}{*}{87.0\small ±0.9} & 86.5 & 85.6 & 86.0 \\
 & N & ~ & 94.0 & 93.6 & 93.8 & ~ & 87.4 & 88.2 & 87.8 \\
 \hline
 \multirow{2}{*}{DDGCN} & R & \multirow{2}{*}{94.8\small ±0.4} & 92.4 & \textbf{97.9} & 95.1 & \multirow{2}{*}{87.8\small ±0.5} & 87.2 & 86.4 & 86.8 \\
 & N & ~ & \textbf{97.6} & 91.7 & 94.6 & ~ & 88.3 & 89.1 & 88.7 \\
 \hline
 \multirow{2}{*}{RAGCL} & R & \multirow{2}{*}{95.9\small ±0.4} & 95.3 & 96.5 & 95.9 & \multirow{2}{*}{89.4\small ±0.4} & 89.4 & 87.7 & 88.5 \\
 & N & ~ & 96.6 & 95.4 & 96.0 & ~ & 89.5 & 90.9 & 90.2 \\
 \hline
 \multirow{2}{*}{\textbf{AD-GSCL(p)}} & R & \multirow{2}{*}{\textbf{97.1\small ±0.2}} & \textbf{96.9} & 97.1 & \textbf{97.0} & \multirow{2}{*}{91.3\small ±0.4} & 91.6 & 89.6 & 90.6 \\
 & N & ~ & 97.2 & \textbf{97.0} & \textbf{97.1} & ~ & 91.2 & 92.9 & 92.0 \\
 \hline
 \multirow{2}{*}{\textbf{AD-GSCL(s)}} & R & \multirow{2}{*}{97.0\small ±0.3} & 96.7 & 97.1 & 96.9 & \multirow{2}{*}{\textbf{92.2\small ±0.3}} & \textbf{92.3} & \textbf{90.7} & \textbf{91.5}\\
 & N & ~ & 97.2 & 96.8 & 97.0 & ~ & \textbf{92.1} & \textbf{93.5} & \textbf{92.8}\\
 \Xhline{1.0pt}
\end{tabular}
\caption{Weibo and DRWeibo experimental results on class-balanced test sets.}
\label{tab:weibo-cbc}
\end{table*}

\begin{table*}[h]
\centering
\begin{tabular}{ccccccccccc}
 \Xhline{1.0pt}
 \rowcolor{gray!20}
 ~ & \multicolumn{5}{c}{\textbf{Twitter15}} & \multicolumn{5}{c}{\textbf{Twitter16}}\\
 \cline{2-11}
 \rowcolor{gray!20}
 ~ & ~ & \textbf{N} & \textbf{F} & \textbf{T} & \textbf{U} & ~ & \textbf{N} & \textbf{F} & \textbf{T} & \textbf{U}\\
 \cline{3-6}
 \cline{8-11}
 \rowcolor{gray!20}
 \multirow{-3}{*}{\textbf{Method}} & \multirow{-2}{*}{\textbf{Acc.}} & \textbf{F1} & \textbf{F1} & \textbf{F1} & \textbf{F1} & \multirow{-2}{*}{\textbf{Acc.}} & \textbf{F1} & \textbf{F1} & \textbf{F1} & \textbf{F1}\\
 \hline
 PLAN & 81.9\small ±0.4 & 83.9 & 85.4 & 81.7 & 75.9 & 84.3\small ±0.5 & \textbf{85.5} & 85.1 & 85.8 & 80.5 \\
 BiGCN & 84.4\small ±0.5 & 85.6 & 84.4 & 86.3 & 80.9 & 88.0\small ±0.9 & 79.3 & 91.2 & 94.7 & 84.9 \\
 GACL & 84.6\small ±0.7 & 85.9 & 84.5 & 86.6 & 81.2 & 89.1\small ±0.4 & 80.2 & 92.9 & 94.5 & 87.2 \\
 DDGCN & 83.5\small ±0.6 & 84.0 & 85.0 & 85.6 & 79.1 & 89.3\small ±0.4 & 80.7 & 93.1 & 94.6 & 87.1 \\
 RAGCL & 85.9\small ±0.5 & 88.3 & 85.9 & 85.1 & \textbf{83.7} & 90.0\small ±0.3 & 83.1 & 91.8 & 95.8 & 87.7 \\
 \textbf{AD-GSCL(p)} & 86.7\small ±0.5 & \textbf{90.2} & 88.3 & 84.6 & 83.1 & 90.8\small ±0.3 & 82.8 & \textbf{93.5} & 95.6 & \textbf{89.7} \\
 \textbf{AD-GSCL(s)} & \textbf{87.4\small ±0.4} & 88.6 & \textbf{89.1} & \textbf{88.1} & 83.6 & \textbf{91.3\small ±0.3} & 84.5 & 92.6 & \textbf{97.9} & 89.3 \\
 \Xhline{1.0pt}
\end{tabular}
\caption{Twitter15 and Twitter16 experimental results on class-balanced test sets.}
\label{tab:twitter-cbc}
\end{table*}

\subsection{Class-Imbalanced Classification}

To validate the efficacy of AD-GSCL in practical applications, we ran class-imbalanced experiments. The class-imbalanced PHEME was selected for the tests. Moreover, to construct class-imbalanced test sets for Weibo and DRWeibo, we expanded the non-rumor class scale by incorporating 3\% of UWeibo data (which is excluded from training) into these two dataset's test sets. The results are shown in Table~\ref{tab:ubc}, including the Area Under the Curve (AUC) and macro F1 score. AD-GSCL outperformed baseline methods, showing enhanced adaptability to real-world scenarios due to our design catering to real-world data domain distribution. AD-GSCL shows a larger improvement over baseline methods in class-imbalanced scenarios compared to class-balanced scenarios. We posit that rumor detection research should utilize datasets reflecting actual data distributions, as it's essential for developing effective models that perform efficiently in real-world environments. From our experiments, we noted that the semi-supervised strategy performs better under class-balanced conditions, while the pre-training strategy proves more effective under class-imbalanced conditions. Therefore, the class distribution within a dataset can significantly guide the selection of the training strategy.

\begin{table}[h]
\centering
\resizebox{0.48\textwidth}{!}{
\begin{tabular}{ccccccc}
 \Xhline{1.0pt}
 \rowcolor{gray!20}
 ~ & \multicolumn{2}{c}{\textbf{Weibo}} & \multicolumn{2}{c}{\textbf{DRWeibo}} & \multicolumn{2}{c}{\textbf{PHEME}}\\
 \cline{2-7}
 \rowcolor{gray!20}
 \multirow{-2}{*}{\textbf{Method}} & \textbf{AUC} & \textbf{F1} & \textbf{AUC} & \textbf{F1} & \textbf{AUC} & \textbf{F1} \\
 \hline
 PLAN & 84.5 & 86.6 & 79.0 & 83.2 & 64.5 & 67.1 \\
 BiGCN & 93.5 & 93.0 & 80.5 & 87.1 & 76.5 & 78.5 \\
 GACL & 89.6 & 91.7 & 82.0 & 86.8 & 74.0 & 78.2 \\
 DDGCN & 90.4 & 93.6 & 79.6 & 86.5 & 77.5 & 79.1 \\
 RAGCL & 91.8 & 92.2 & 79.4 & 83.6 & 74.4 & 78.5 \\
 \textbf{AD-GSCL(p)} & \textbf{96.7} & 95.6 & \textbf{90.6} & \textbf{92.1} & \textbf{78.1} & \textbf{82.3} \\
 \textbf{AD-GSCL(s)} & 96.2 & \textbf{96.0} & 89.6 & 91.7 & 78.0 & 81.8 \\
 \Xhline{1.0pt}
\end{tabular}
}
\caption{Results on class-imbalanced test sets.}
\label{tab:ubc}
\end{table}

\subsection{Ablation Study}

\begin{table}[h]
\centering
\resizebox{0.48\textwidth}{!}{
\begin{tabular}{ccccc}
 \Xhline{1.0pt}
 \rowcolor{gray!20}
 ~ & ~ & ~  & \multicolumn{2}{c}{\textbf{Dataset}} \\
 \cline{4-5}
 \rowcolor{gray!20}
 \multirow{-2}{*}{$\mathcal{L}_{scl}$} & \multirow{-2}{*}{$\mathcal{L}_{ucl}$} & \multirow{-2}{*}{\textbf{Unlabeled Data}} & \textbf{Weibo} & \textbf{DRWeibo} \\
 \hline
  & & & 93.2\small ±0.6 & 86.5\small ±0.7\\
 \checkmark & & & 95.2\small ±0.7 & 89.8\small ±0.6\\
 \checkmark & & \checkmark & 96.1\small ±0.5 & 91.1\small ±0.4\\
 \checkmark &  \checkmark & \checkmark & \textbf{97.0\small ±0.3} & \textbf{92.2\small ±0.3}\\
 \Xhline{1.0pt}
\end{tabular}
}
\caption{The influence of loss and unlabeled data.}
\label{tab:ab}
\end{table}

Some ablation studies have been performed to examine the role of various loss components and unlabeled data, as shown in Table~\ref{tab:ab}. Experiments were conducted under class-balanced condition and semi-supervised strategy. The results show that the supervised contrastive loss and unlabeled data can improve the performance of GNNs. The unsupervised contrastive loss can further improve performance by making massive non-rumor class claims serve as better anchors. These results demonstrate that the modifications made to existing graph contrastive losses in AD-GSCL make it better suited for the domain distribution in rumor detection.

\subsection{Few-shot Rumor Detection}

\begin{figure}[t]
  \centering
  \includegraphics[width=0.48\textwidth]{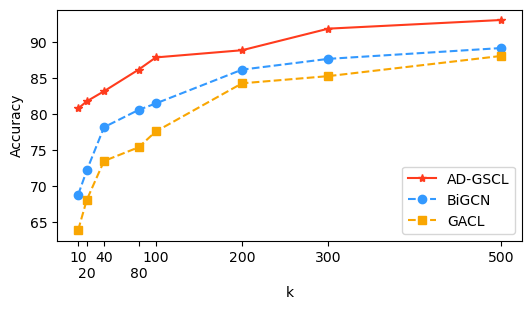}
  \caption{Results of few-shot experiments.}
  \label{fig:fs}
\end{figure}

In a class-balanced setting, we tested BiGCN, GACL, and AD-GSCL on Weibo dataset for few-shot learning (Figure~\ref{fig:fs}). As rumors are typically removed once detected, the accumulation of a large-scale labeled dataset becomes challenging, thereby rendering the exploration of rumor detection in few-shot settings as highly significant. Our experiments were conducted with varying numbers of labeled samples, denoted as $k$, including 10, 20, 40, 80, 100, 200, 300, and 500. An initial pre-training phase was conducted on UWeibo, followed by fine-tuning on the labeled samples. The results suggest that, even with a few labeled samples, AD-GSCL can effectively utilize unlabeled data to learn superior claim representations. This leads to appreciable performance, reinforcing the potential utility of AD-GSCL in practical, data-scarce scenarios.

\section{Conclusion}

In this study, we analyzed the distribution characteristics of rumor detection datasets and real scenarios. 
We introduced AD-GSCL, adapting graph contrastive learning for rumor detection. 
Our extensive experiments demonstrated AD-GSCL’s superiority over baselines. 
The proposed paradigm provides an impactful approach and modeling strategies for handling imbalanced distributions.

\section*{Ethical Statement}

In this research, we place a high emphasis on ethical considerations. We utilized web crawling tools to collect data from publicly posted content by users on the Weibo and Twitter platforms, which is visible to any user of these platforms. We will process the final public data by removing personally identifiable information to ensure that no individual can be identified from our dataset. We explicitly state that the sole purpose of collecting and analyzing data is for academic research, aimed at improving the quality of information on social media and reducing the spread of rumors. By employing semi-supervised learning methods, we are able to enhance the performance of models under conditions of scarce labeled data, which holds significant academic and social value for rumor detection research. We commit to adhering to ethical principles throughout the research process, ensuring that the conduct of the research complies with legal requirements and moral standards, and treating our data, participants, and society with responsibility.

\section*{Limitations}

AD-GSCL is a semi-supervised method that applies large-scale unlabeled datasets. Compared with the existing methods based on propagation structure learning, it needs to process more large-scale data, so it will occupy more computing resources.
Furthermore, our research only investigated the distribution of data on social media platforms during typical periods. However, in certain countries or regions during special circumstances (such as pandemics, elections, or wars), the distribution of data may change.

\section*{Acknowledgments}

The authors would like to thank all the anonymous reviewers for their help and insightful comments.
This work is supported in part by the National Key R\&D Program of China (2018AAA0100302) and the National Natural Science Foundation of China (61876016).

\bibliography{custom}

\appendix

\section{Dataset Construction and Statistics}

In this section, we will elaborate on the construction of large-scale unlabeled datasets from the Weibo and Twitter platforms, as well as the methodology employed for the statistical analysis of domain distributions within the datasets. Additionally, we will supplement the domain distribution statistics for UTwitter, Twitter15, and Twitter16 datasets.

\subsection{Claim Propagation Trees}
\label{sec:rpt}

\begin{figure}[h]
  \centering
  \subfigure[Rumor]{\includegraphics[width=0.48\linewidth]{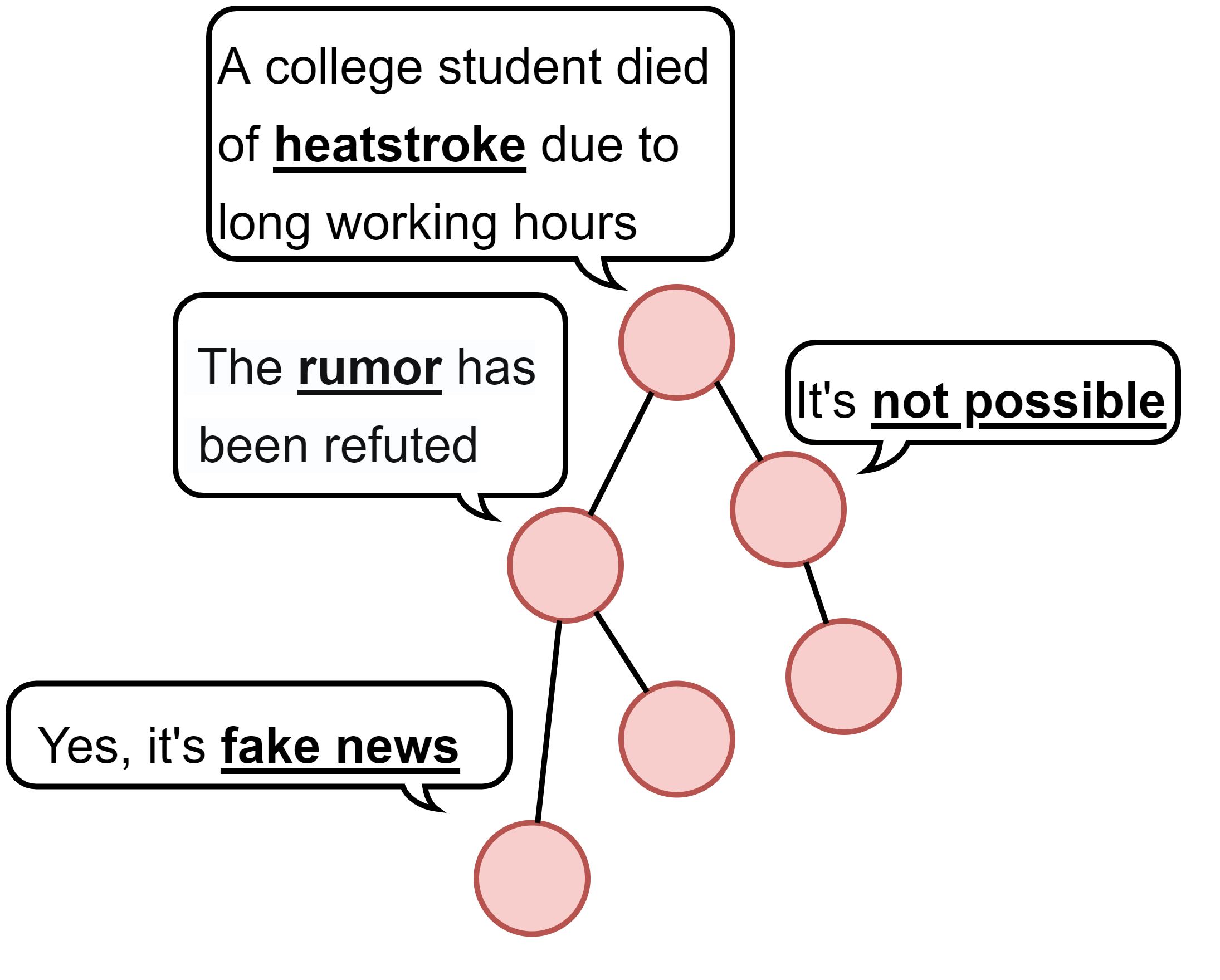}}
  \subfigure[Non-rumor]{\includegraphics[width=0.48\linewidth]{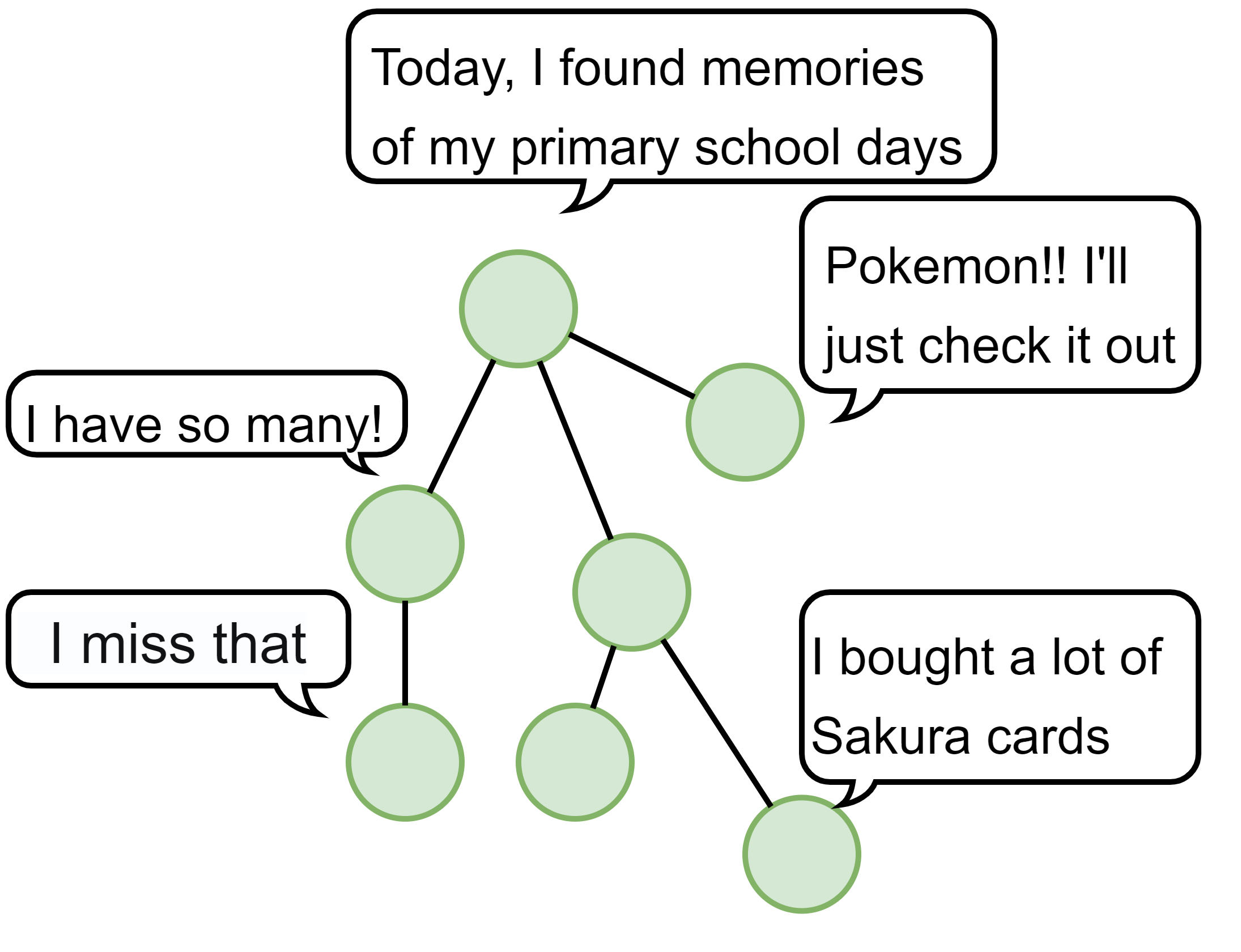}}
  \caption{Examples of propagation trees. Comments under rumor thread typically express more heated stances.}
  \label{fig:pt}
\end{figure}

Figure~\ref{fig:pt} illustrates two examples of claim propagation trees sourced from the Twitter platform. In these examples, the responses to rumor claims and non-rumor claims reveal notable differences in both stance and sentiment. These characteristics are crucial for effective rumor identification.

\subsection{Unlabeled Dataset Construction}
\label{sec:udc}

For the UWeibo dataset, we employed web crawler techniques to randomly collect trending posts and their complete propagation structures from the homepage of popular Weibo posts\footnote{\url{https://weibo.com/hot/weibo/102803}}. To ensure the dataset's integrity and independence from platform recommendation algorithms, we utilized multiple newly created accounts to extract data. This approach aimed to mitigate potential biases that might arise from the platform's algorithms and to reflect the genuine domain distribution of social media content. Regarding UTwitter dataset, we initially utilized multiple newly created accounts to randomly follow high-follower count influencers. Subsequently, we conducted random crawling of posts and their propagation structures from the Twitter homepage\footnote{\url{https://twitter.com/home}}. Due to the fact that UTwitter dataset is exclusively sourced from users with a substantial number of followers, the authenticity of the posts is more likely to be ensured. The data in the UTwitter dataset spans from October 2022 to April 2023, while the data in the UWeibo dataset covers from March 2022 to April 2023. The data is uniformly distributed over time. The code for the web scraping program can be found at \url{https://github.com/CcQunResearch/WeiboPostAndCommentCrawl} and \url{https://github.com/CcQunResearch/TwitterPostAndCommnetCrawl}.

Due to the stringent regulation imposed by platforms on the dissemination of rumors, acquiring a sufficiently large-scale labeled dataset for rumor detection proves to be exceptionally challenging. Conversely, obtaining extensive amounts of unlabeled data is relatively simpler, especially with the availability of platform data APIs offered by certain mainstream social media platforms (e.g., Twitter API). Consequently, we posit that future research should place greater emphasis on semi-supervised rumor detection methods.

\subsection{Domain Statistical Methodology}
\label{sec:dsm}

For the Twitter platform data, we manually performed the statistical analysis of its distribution. For UTwitter dataset, we specifically examined the domain distribution of a subset comprising 1000 data instances. Additionally, we analyzed the section distribution of the union of Twitter15 and Twitter16 datasets. Given that these two datasets operate at the event level\cite{t1516rep}, with multiple claims associated with the same event, we implemented a deduplication process based on overlapping claims and events found within the two datasets.

\begin{figure*}[h]
  \centering
  \subfigure[Domain]{\includegraphics[width=0.766\textwidth]{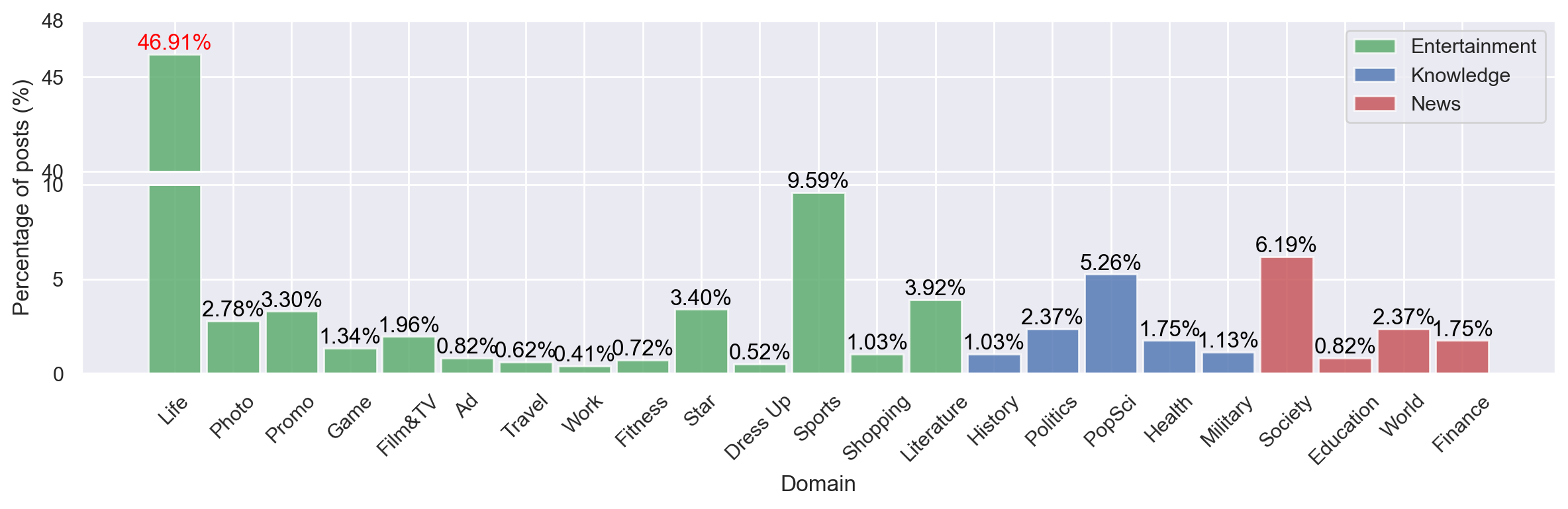}}
  \subfigure[Section]{\includegraphics[width=0.184\textwidth]{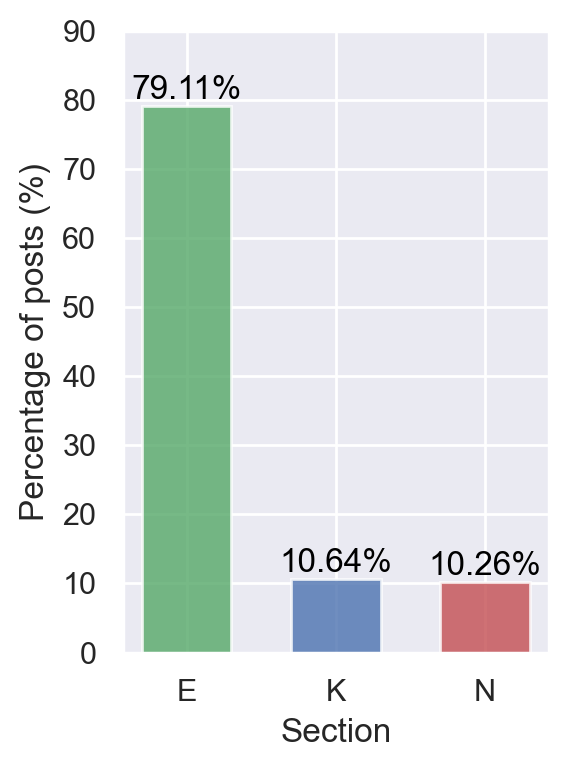}}
  \caption{Distribution of post counts by domain and section on UTwitter dataset. E, K and N represent the three sections Entertainment, Knowledge and News respectively.}
  \label{fig:ds_utwitter}
\end{figure*}

\begin{figure*}[h]
  \centering
  \subfigure[Non-Rumor]{\includegraphics[width=0.184\textwidth]{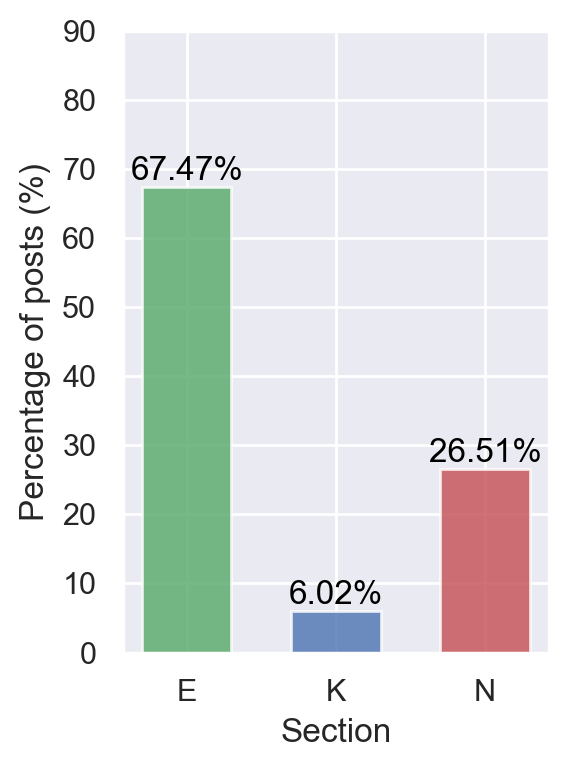}}
  \subfigure[True Rumor]{\includegraphics[width=0.184\textwidth]{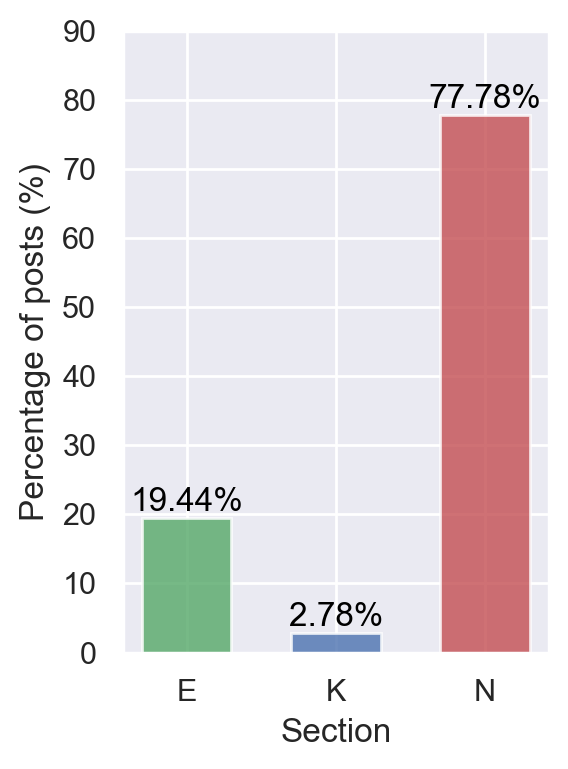}}
  \subfigure[False Rumor]{\includegraphics[width=0.184\textwidth]{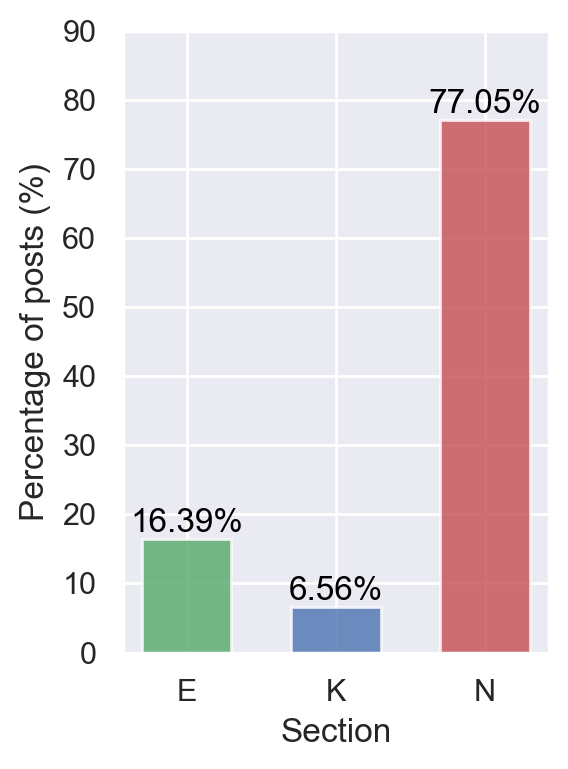}}
  \subfigure[Unverified Rumor]{\includegraphics[width=0.184\textwidth]{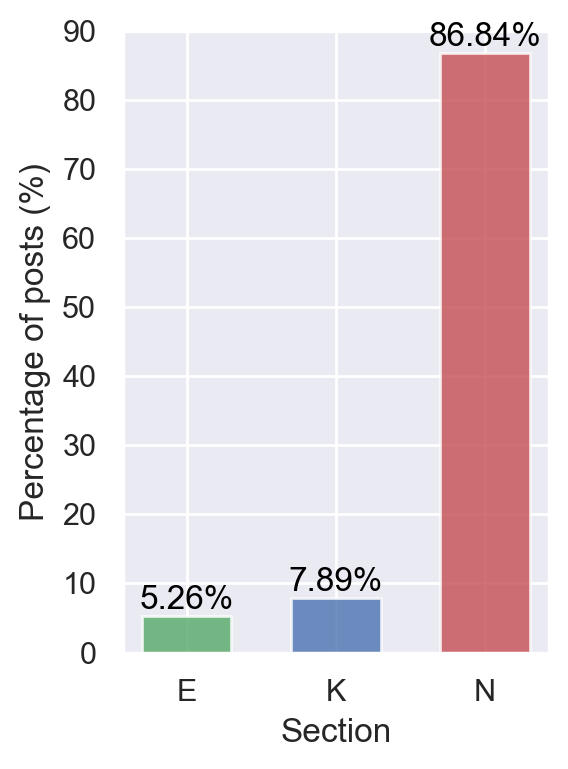}}
  \caption{Post section distribution on Twitter15 and Twitter16 dataset.}
  \label{fig:section_twitter}
\end{figure*}

The domain and section distributions for the UTwitter dataset, as well as the section distribution for the union of Twitter15 and Twitter16 datasets, are illustrated in Figure~\ref{fig:ds_utwitter} and Figure~\ref{fig:section_twitter}, respectively. Specifically, similar to the domain distribution observed on Weibo platform, the following trends were identified: (1) As for UTwitter Dataset, the majority of posts on the Twitter platform belong to Entertainment section (79.11\%), with a substantial portion consisting of personal life-sharing content (46.91\%). (2) The section distribution between non-rumor and rumor data in the labeled Twitter15 and Twitter16 datasets shows significant differences. Specifically, the posts in all three Rumor classes are predominantly associated with News section (77.78\%, 77.05\%, 86.84\%). (3) The section distribution of the unlabeled data in UTwitter exhibits striking similarity to the section distribution of non-rumor data in Twitter15 and Twitter16. In other words, the majority of posts in both cases belong to Entertainment section that is less likely to generate rumors or misinformation. These statistical findings further validate the rationale behind utilizing large-scale unlabeled data as non-rumor class instances and provide an explanation for why it is unreasonable to classify them as other Rumor classes.

Indeed, the analysis of unlabeled data from both Weibo and Twitter platforms reveals some subtle differences. For instance, the presence of sports-related posts is relatively lower on the Weibo platform compared to the Twitter platform, while history-related posts are more prevalent on Weibo but less so on Twitter. These disparities could potentially be attributed to cultural differences between the Eastern and Western regions. Cultural variations play a significant role in shaping the content and user preferences on different social media platforms. The observed differences in content themes might reflect the diverse interests and priorities of the respective user bases. Factors such as language, geographic location, historical background, and societal norms can influence the types of content shared and engaged with on social media platforms.

\section{Experiment Details}
\label{sec:ed}

In this section, we mainly introduce some specific settings in our experiments, including the way of data preprocessing and the hyperparameter configuration when training the model.

\subsection{Data Preprocessing}

For the texts in all datasets, we first standardize the different fonts present in the texts, then identify user mentions and web/url links as special tokens, \texttt{<@user>} and \texttt{<url>}. Next, we use the TweetTokenizer from the NLTK toolkit and jieba word segmentation engine to tokenize the raw texts in English and Chinese datasets, respectively. Additionally, we use the \texttt{emoji} package\footnote{\url{https://pypi.org/project/emoji}} to translate the emojis in the texts into text string tokens.

\subsection{Hyperparameter Configuration}

All models are implemented by PyTorch and the baseline methods are re-implemented. GACL uses BERT \cite{bert} to extract the initial feature vector of each post in the propagation tree. In addition to GACL, other models use 200-dimensional word2vec word embeddings \cite{word2vec} as initial feature vectors. 

We set the controlling hyperparameters for the loss functions $\lambda$, $\alpha$, and $\beta$ uniformly to a small value of 1e-3. The $\tau$ of supervised contrastive loss is set to 0.3, while the batch size is set to 32, and the learning rate is set to 1e-3. The GCN encoder consists of 3 layers, and we employ sum-pooling for node representations to obtain graph representation. We optimize the loss function using the Adam optimizer \cite{adam}. The entire training process for all models is conducted on a single Nvidia GeForce RTX 3090 GPU.

Note that BiGCN and GACL employ early stopping to observe the performance attainable by the models. However, due to oscillations during the early stages of model training, the observed model performance becomes unstable. To ensure a fairer comparison of different models' performance, we conduct experiments using the same dataset on AD-GSCL and multiple baseline methods. Furthermore, all models are trained for 100 epochs until convergence. We consider the average results of the last 10 epochs among these 100 epochs as the stable results achievable by the models. This approach helps mitigate the effects of training fluctuations and provides a more reliable basis for comparing the performance of the models.

\section{Supplementary Experiments}
\label{sec:se}

We conducted a series of additional ablation experiments to delve into the impact of different modules in AD-GSCL more comprehensively. Additionally, we extended the few-shot experiments on DRWeibo dataset. 

\subsection{Extended Ablation Study}

We conducted a series of extended ablation experiments to verify the influence of different factors on the model performance. Experiments were conducted under class-balanced condition and semi-supervised training strategy.

\subsubsection{Adaptation Measures} 

\begin{table}[t]
\centering
\begin{tabular}{lcc}
 \Xhline{1.0pt}
 \rowcolor{gray!20}
 ~ & \textbf{Weibo} & \textbf{DRWeibo}\\
 \hline
 AD-GSCL & 97.0\small ±0.3 & 92.2\small ±0.3 \\
 \hdashline
 \quad w/ Standard SCL & 95.4\small ±0.6 & 89.4\small ±0.5 \\
 \quad w/ Standard MI Max. & 95.9\small ±0.4 & 90.5\small ±0.4 \\
 \Xhline{1.0pt}
\end{tabular}
\caption{The influence of adaptation measures.}
\label{tab:am}
\end{table}

In order to enable the supervised contrastive learning method and the unsupervised contrastive learning method of MI maximization adopted by AD-GSCL to adapt to the distribution and structural characteristics of social media data, we have made the following designs respectively: (1) Modified the process of constructing positive and negative samples in standard supervised contrastive learning \cite{scl}, restricting non-rumor data from participating in positive sample construction; (2) Changed the standard MI maximization contrastive loss (maximizing MI between node representations and graph representations) to maximizing the MI between node representations and root node representations. We explored the effects of these two designs in Table~\ref{tab:am}. Experimental results on Weibo and DRWeibo showed that both designs improved model performance to some extent. This indicates that for existing contrastive learning methods, adapting them according to the characteristics of social media data when applying them to rumor detection tasks is very important.

\subsubsection{Graph Directionality}

\begin{figure}[t]
  \centering
  \subfigure[Weibo]{\includegraphics[width=0.23\textwidth]{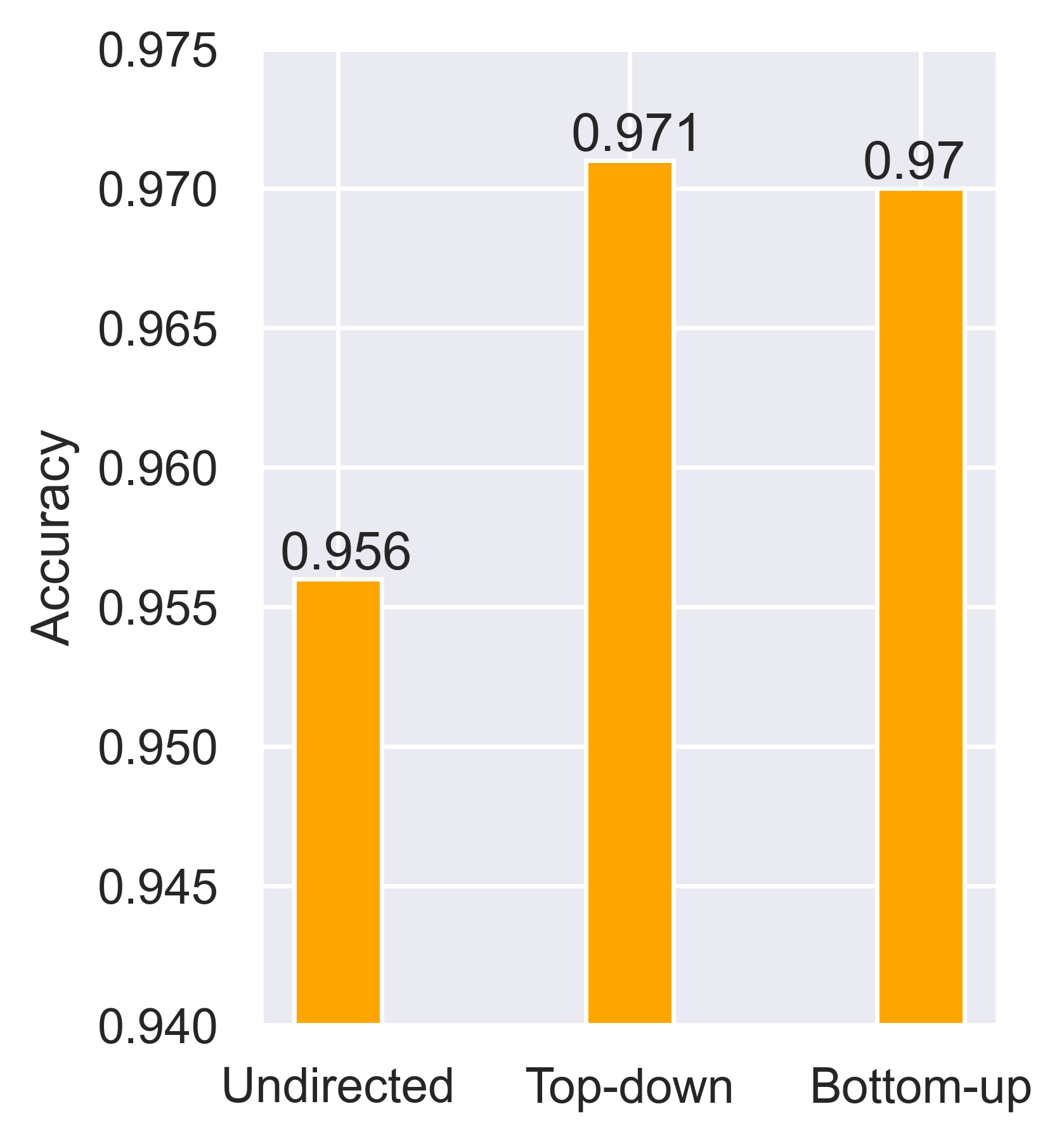}}
  \subfigure[DRWeibo]{\includegraphics[width=0.23\textwidth]{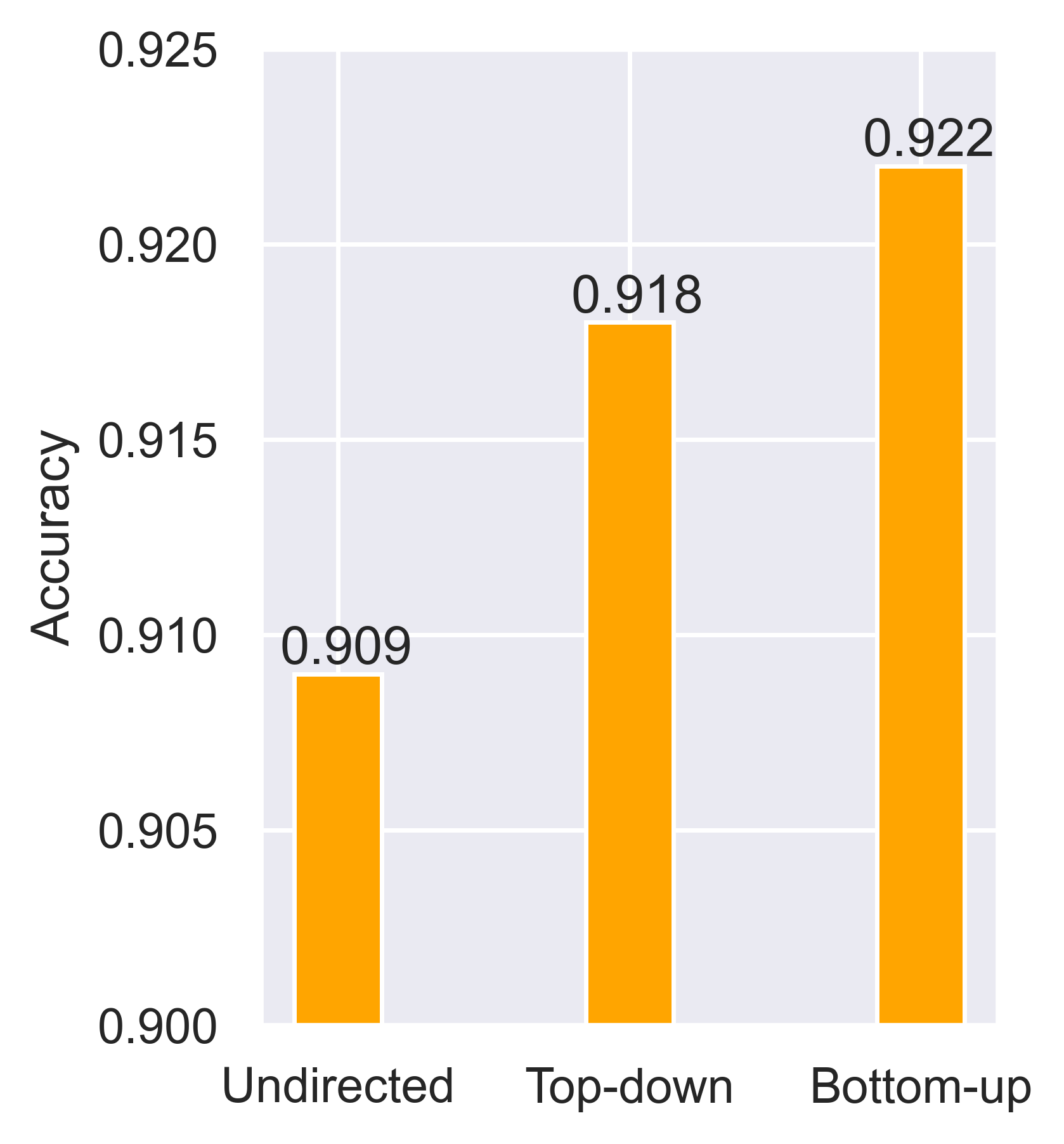}}
  \caption{The influence of information flow direction.}
  \label{fig:direction}
\end{figure}

In Figure~\ref{fig:direction}, we explored the effects of different graph directionality on performance on Weibo and DRWeibo datasets. The results show that undirected graphs lead to performance degradation. This may be because in the forward propagation of GNNs, nodes densely connected at the root node can see each other in their neighborhood views, aggregating each other's information and eventually losing the uniqueness of node features, causing over-smoothing. In contrast, top-down and bottom-up directed graphs can effectively block excessive information flow between nodes at the root. For AD-GSCL, since we use the root node representation as the global representation in unsupervised training, we adopt a bottom-up directed graph, which allows the root node to aggregate features from most nodes in the graph.

\subsubsection{Mutual Information Estimators}

\begin{table}[t]
\centering
\begin{tabular}{ccc}
 \Xhline{1.0pt}
 \rowcolor{gray!20}
 \textbf{Measures} & \textbf{Weibo} & \textbf{DRWeibo}\\
 \hline
 JS & 97.0\small ±0.3 & 92.2\small ±0.3 \\
 KL & 96.6\small ±0.6 & 92.0\small ±0.4 \\
 DV & 97.1\small ±0.3 & 91.6\small ±0.5 \\
 \Xhline{1.0pt}
\end{tabular}
\caption{The influence of MI estimators.}
\label{tab:mie}
\end{table}

We investigated the effects of MI measures in unsupervised training on Weibo and DRWeibo datasets, with experimental results shown in Table~\ref{tab:mie}. The results demonstrate that AD-GSCL is relatively robust to the choice of MI estimator. We select the Jensen-Shannon MI estimator \cite{js} which gives the most stable performance.

\subsection{Extended Few-shot Experiments}

\begin{figure}[t]
  \centering
  \includegraphics[width=0.48\textwidth]{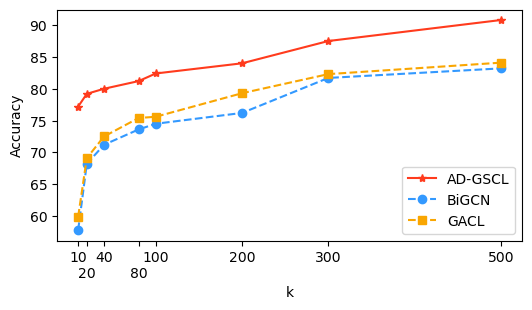}
  \caption{Results of few-shot experiments.}
  \label{fig:fsdr}
\end{figure}

We conducted few-shot experiments on DRWeibo dataset with the same settings as the Weibo dataset. The experimental results are shown in Figure~\ref{fig:fsdr}. Similar results are observed on DRWeibo as on Weibo dataset, that is, the fewer labeled samples, the more significant improvement of AD-GSCL over baselines. This demonstrates the superiority of semi-supervised methods under few-shot conditions.

\end{document}